\documentstyle[epsfig,epsf,indentfirst,bm,12pt]{article}
\begin{document}

\author{F.-K. Guo$^{1,6}$\footnote{\tt{e-mail: guofk@mail.ihep.ac.cn}},
P.-N. Shen$^{2,1,4,5}$, H.-C. Chiang$^{3,1}$,
R.-G. Ping$^{2,1}$\\
{\small $^1$Institute of High Energy Physics, Chinese Academy of Sciences,}\\
{\small P.O.Box 918(4), Beijing 100049, China}\\
{\small $^2$CCAST(World Lab.), P.O.Box 8730, Beijing 100080, China}\\
{\small $^3$South-west Normal University, Chongqing 400715, China}\\
{\small $^4$Institute of Theoretical Physics, Chinese Academy of
Sciences, P.O.Box 2735, China}\\
{\small $^5$Center of Theoretical Nuclear Physics, National
Laboratory of Heavy Ion Accelerator,}\\
{\small Lanzhou 730000, China}\\
{\small $^6$Graduate School of the Chinese Academy of Sciences,
Beijing 100049, China}}

\date{}

\title{\Large{\bf{Heavy Quarkonium $\pi^+\pi^-$ Transitions and
a Possible $b\bar{b}q\bar{q}$ State}}}

\maketitle

\begin{abstract}
$\pi^+\pi^-$ transitions of heavy quarkonia, especially the $%
\Upsilon(3S)\to\Upsilon(1S)\pi^+\pi^-$ decay process, are
revisited. In the framework of the Chiral Unitary Theory (ChUT),
the $S$ wave $\pi\pi$ final state interaction (FSI) is included.
It is found that when an additional
intermediate state with $J^P=1^+$ and $I=1$ is introduced, not only the $%
\pi\pi$ invariant mass spectrum and the $cos\theta_\pi^*$ distribution in
the $\Upsilon(3S)\to\Upsilon(1S)\pi^+\pi^-$ process can simultaneously be
well-explained, but also a consistent description for other bottomonia $%
\pi^+\pi^-$ transitions can be obtained. As a consequence, the mass and the
width of the intermediate state are predicted. From the quark content
analysis, this state should be a $b\bar bq\bar q$ state.
\end{abstract}

\vspace{0.6cm}

%{\bf Keywords:}

\vspace{1.5cm}

\section{Introduction}

In recent years, more and more decay data of heavy quarkonia have been
accumulated, and new information on hadron physics has been extracted. Many
investigations along this line, for instance, the decay properties of the $%
\Psi(2S)\to J/\Psi\pi^+\pi^-$, $\Upsilon(2S)\to\Upsilon(1S)\pi^+\pi^-$, $%
\Upsilon(3S)\to\Upsilon(2S)\pi^+\pi^-$, and $\Upsilon(3S)\to\Upsilon(1S)%
\pi^+\pi^-$ processes have been carried out. A commonly used method for such
studies is the QCD multipole expansion method proposed by Gottfried \cite%
{gott78} and further developed by others \cite%
{yan80,vz80,ns81,ky81,mo89,zk91}. It was shown that although most
data of the mentioned processes can be well-reproduced, the
$\pi\pi$ invariant mass
spectrum and the angular distribution of the $\Upsilon(3S)\to\Upsilon(1S)%
\pi^+\pi^-$ decay cannot satisfactorily be described.

Phenomenological models \cite{lt88,bc75,mu97,ywz99,ue03} are also
used in such studies. For instance, in Ref. \cite{mu97}, Mannel et
al. constructed an effective Lagrangian on the basis of the chiral
perturbation theory (ChPT) and the heavy quark non-relativistic
expansion. Under the approximation in the limits of the chiral
symmetry and the heavy quark mass \cite {ckk94}, the measured
$\pi\pi$ invariant mass spectra of the $\Psi(2S)\to
J/\Psi\pi^+\pi^-$, $\Upsilon(3S)\to\Upsilon(2S)\pi^+\pi^-$ and $
\Upsilon(2S)\to\Upsilon(1S)\pi^+\pi^-$ decay processes were very
well fitted, but not of the decay
$\Upsilon(3S)\to\Upsilon(1S)\pi^+\pi^-$. In order to explain the
data of $d\Gamma(\Upsilon(3S)\to
\Upsilon(1S)\pi^+\pi^-)/dm_{\pi\pi}$, the $\pi\pi$ S-wave FSI was
included by using a parameterization. However, although the coupling
constant ratios $(g_2/g_1)_{b \bar{b}}$ in the
$\Upsilon(2S)\to\Upsilon(1S)\pi^+\pi^-$ and
$\Upsilon(3S)\to\Upsilon(2S)\pi^+\pi^-$ decays are approximately
equal to each other, they are much smaller than that in the
$\Upsilon(3S)\to\Upsilon(1S)\pi^+\pi^-$ decay. The later one is
about 10 times larger than the former. It seems unnatural. Moreover,
M.-L. Yan et al. \cite{ywz99} pointed out that the parametrization
of the S-wave FSI there was not properly carried out because in the
$g_2$ term in the amplitude, there are also $D$ wave components. In
Ref.~\cite{ue03}, according to a unitarized chiral theory, an $S$
wave effective Lagrangian and an effective scalar form factor were
adopted. As a result, the $\pi\pi$ invariant mass spectra of heavy
quarkonium decays can be reproduced, but the $\cos\theta_\pi^*$
distribution still cannot be explained.

In the $\Upsilon(3S)\to\Upsilon(1S)\pi^+\pi^-$ decay, the two-peak
structure of the $\pi^+\pi^-$ invariant mass spectrum has been
considered as a consequence of $\pi\pi$ FSI
\cite{bdm89,ck93,ckk94,gmp04,lr02} or the additional contribution
from the $D$ wave component of $\Upsilon(3S)$ \cite{ckk93}.
Ignoring the contribution from the higher order pion momentum,
Chakravarty {\it et al.} \cite{ckk94} explained the $\pi\pi$
invariant mass spectrum of $\Upsilon(3S)\to\Upsilon(1S)\pi^+\pi^-$
with $\chi^2/N_{d.f.}$=11.0/7 or C.L.=14.0\%, but only
quantitatively gave the $\cos\theta_\pi^*$ distribution. Gallegos
et al. \cite{gmp04} parametrized a more generalized amplitude in
which both $S$ and $D$ wave contributions are included by fitting
to the invariant mass distributions and the angular distributions
of the decays mentioned at the beginning of this section. It is
clear that the result of the parametrization should be consistent
with the existing $\pi\pi$ scattering data. Whether this condition
has been satisfied in that calculation remains a question.
L\"{a}hde et al. \cite{lr02} showed that in order to better
describe the $\pi\pi$ invariant mass spectra, the contribution of
the pion re-scattering should be small in the $\Psi(2S)\to
J/\Psi\pi^+\pi^-$ or $\Upsilon(2S)\to \Upsilon(1S)\pi^+\pi^-$
decay, but dominant in the $\Upsilon(3S)\to\Upsilon(1S)\pi^+\pi^-$
decay. Why it is so is still a puzzle.

On the other hand, by introducing a $b\bar{b}q\bar{q}$ resonance
with $J^P=1^+$ and mass of 10.4-10.8 GeV, Anisovich et al.
\cite{absz95} explained the $\pi^+ \pi^-$ invariant mass spectra
of above mentioned bottomonium decays, but not the angular
distribution of $\Upsilon(3S)\to\Upsilon(1S)\pi^+\pi^-$.

To treat meson-meson $S$ wave FSI properly, a chiral
nonperturbative approach, called chiral unitary theory (ChUT), was
recently proposed by Oller and Oset \cite{oo97} and later
developed by themselves and others
\cite{oop98,oo98,oond,ol98,mo01,ue03,lov04,prov03,rpoc04,lo98,oo01,ol05}
(for details, refer to the review article Ref.~\cite{oor00}). In
this theory \cite{oo97}, the coupled channel Bethe-Salpeter
equations (BSE), in which the lowest order amplitudes in the ChPT
are employed as the kernel, are solved to resum the contributions
from the s-channel loops of the re-scattering between pseudoscalar
mesons. By properly choosing the three-momentum cutoff, the only
free parameter in the theory, ChUT can well-describe the data of
the $S$ wave meson-meson interaction up to $\sqrt{s}\simeq1.2GeV$
\cite{oo97} which is much larger than the energy in the region
where the standard ChPT is still valid, and can dynamically
generate the scalar resonances $\sigma$ ($f_0(600)$), $ f_0(980)$
and $a_0(980)$ \cite{oo97}.

In this work, we adopt an amplitude used in Ref. \cite{mu97},
which includes both $S$ and $D$ wave contributions, and consider
the $S$ wave $\pi\pi$ FSI in the framework of ChUT to study the
$\Psi(2S)\to J/\Psi\pi^+\pi^-$,
$\Upsilon(2S)\to\Upsilon(1S)\pi^+\pi^-$,
$\Upsilon(3S)\to\Upsilon(2S)\pi^+\pi^-$ and
$\Upsilon(3S)\to\Upsilon(1S)\pi^+\pi^-$ decays.

The paper is organized as follows: In Section~\ref{sec:for}, the effective
Lagrangian and ChUT are briefly introduced. In terms of the $t$-matrix
written in Section~\ref{sec:for}, the $\Psi(2S)\to J/\Psi\pi^+\pi^-$ decay
is discussed in Section~\ref{sec:psi}. Section~\ref{sec:upsilon} is
dedicated to the bottomonium $\pi^+ \pi^-$ transitions, and a brief summary
is given in Section~\ref{sec:sum}.

\section{Brief formalism}

\label{sec:for} In the heavy quarkonium $\pi^+ \pi^-$ transition process,
the Lagrangian in the lowest order, which appropriately incorporates the
chiral expansion with the heavy quark expansion, can be written as \cite%
{mu97}
\begin{equation}
{\cal L}= {\cal L}_0+{\cal L}_{S.B.}
\end{equation}
\begin{eqnarray}
{\cal L}_0 &\!\!\!\!=&\!\!\!\!
g_1A_{\mu}^{(v)}B^{(v)\mu*}Tr[(\partial_{\nu}U)(\partial^{\nu}U)^\dag]
+g_2A_{\mu}^{(v)}B^{(v)\mu*}Tr[(v\cdot{\partial}U)(v\cdot{\partial}U)^\dag]
\nonumber \\
&\!\!\!\!+&\!\!\!\!
g_3A_{\mu}^{(v)}B_{\nu}^{(v)*}Tr[(\partial^{\mu}U)(\partial^{\nu}U)^\dag
+(\partial^{\mu}U)^\dag(\partial^{\nu}U)]+h.c.
\end{eqnarray}
\begin{eqnarray}
{\cal L}_{S.B.} &\!\!\!\!=&\!\!\!\!
g_4A_{\mu}^{(v)}B^{(v)\mu*}Tr[M(U+U^{\dag}-2)]  \nonumber \\
&\!\!\!\!+&\!\!\!\!
ig_{^{\prime}}\varepsilon^{\mu\nu\alpha\beta}[v_{\mu}A_{\nu}^{(v)}\partial_{%
\alpha}B_{\beta}^{(v)*}
-(\partial_{\mu}A_{\nu}^{(v)})v_{\alpha}B_{\beta}^{(v)*}]Tr[M(U-U^{%
\dag})]+h.c.
\end{eqnarray}
where $g_i$ denotes the coupling constants, $U$ is a $3{\times}3$
matrix that contains the pseudoscalar Goldstone fields, and $M=
diag\{m_u, m_d, m_s\}$ is the quark mass matrix with $m_u$, $m_d$,
$m_s$ being the masses of current quarks $u$, $d$ and $s$,
respectively. $A_\mu^{(v)}$ and $B_\mu^{(v)}$ are the fields of
the initial and final states of heavy vector quarkonia,
respectively, and $v$ is the velocity vector of $A$. The tree
diagram amplitude for the decay of a vector meson into two
pseudoscalar mesons and one vector meson in the rest frame of
decaying particle can be expressed as \cite{ck93,ckk94,mu97}
\begin{equation}  \label{eq:mannel}
t=-\frac{4}{f^{2}_\pi}[(g_1p_1\cdot p_2+g_{2}p_1^0p_2^0
+g_3m_\pi^2)\varepsilon^*\cdot\varepsilon^{^{\prime}}+g_4(p_{1\mu}p_{2%
\nu}+p_{1\nu}p_{2\mu}) \varepsilon^{*\mu}\varepsilon^{^{\prime}\nu}]
\end{equation}
where $f_\pi=93MeV$ is the decay constant of pion, $p_1$ and $p_2$
are the four-momenta of $\pi^+$ and $\pi^-$, respectively, $p_1^0$
and $p_2^0$ denote the energies of $\pi^+$ and $\pi^-$ in the lab
frame, and $\varepsilon $ and $\varepsilon^{^{\prime}}$ are the
polarization vectors of the heavy quarkonia, respectively. It can
be verified by the CLEO data \cite{ckk94} that by considering the
chiral symmetry breaking scale and the heavy quark mass, the
contribution from the last term ($g_4$-term) is strongly
suppressed \cite{mu97}. Thus, the $g_4$-term in
Eq.~(\ref{eq:mannel}) can be ignored and the amplitude can further
be written as
\begin{eqnarray}  \label{eq:v0}
V_{0}=-\frac{4}{f^{2}_\pi}(g_1p_1\cdot p_2+g_{2}p_1^0p_2^0
+g_3m_\pi^2)\varepsilon^*\cdot\varepsilon^{^{\prime}}.
\end{eqnarray}
It is noted that the $D$ wave component exists in the $g_2$ term \cite{bc75,
ywz99}. Under Lorentz transformation, $p_1^0$ and $p_2^0$ can be expressed
as the functions of the momenta of pions in the center of mass (c.m.) frame
of the $\pi\pi$ system:
\begin{eqnarray}
p_1^0 &\!\!\!\!=&\!\!\!\! \frac{1}{\sqrt{1-{\bm\beta}^2}}(p_1^{0*}+|{\bm %
\beta}||{\mathbf p}_1^{*}|\cos\theta_{\pi}^*), \\
p_2^0 &\!\!\!\!=&\!\!\!\! \frac{1}{\sqrt{1-{\bm\beta}^2}}(p_1^{0*}-|{\bm \beta%
}||{\mathbf p}_1^{*}|\cos\theta_{\pi}^*),
\end{eqnarray}
where ${\bm\beta}$ is the velocity of the $\pi\pi$ system in the
rest frame of the initial particle, $p_1^*$=($p_1^{0*}$, ${\mathbf
p}_1^*$) and $p_2^* $=($p_2^{0*}$, ${\mathbf p}_2^*$) are the
four-momenta of $\pi^+$ and $\pi^-$ in the c.m. frame of the
$\pi\pi$ system, respectively. So $p_1^0p_2^0$ can be decomposed
as
\begin{eqnarray}
p_1^0p_2^0=\frac{1}{1-{\bm \beta}^2}[(p_1^{0*2}-\frac{{\bm \beta}^2{\mathbf p}_1^{*2}%
}{3})P_0(\cos\theta_{\pi}^*)-2{\bm \beta}^2{\mathbf
p}_1^{*2}P_2(\cos\theta_{\pi}^*)]
\end{eqnarray}
where $P_0(\cos\theta_{\pi}^*)=1$ and
$P_2(\cos\theta_{\pi}^*)=\frac{1}{2}(\cos^2\theta_{\pi}^*-\frac{1}{3})$
are the Legendre functions of the 0-th order and 2-nd order,
respectively.

Furthermore, the $S$ wave $\pi \pi$ FSI which is important in this
energy region should properly be included into the theoretical
calculation. ChUT \cite{oo97} is one of the suitable approaches
for this job, because by using this theory, the $S$ wave $\pi -\pi
$ scattering data up to 1.2GeV can be well reproduced. However,
ChPT amplitudes in the $O(p^{2})$ order are adopted as the kernel
of the coupled-channel BSE \cite{oo97}, so the $D$ wave FSI cannot
be included. In the decays considered, the kinematical region is
below $0.9GeV$ for the decay
$\Upsilon(3S)\to\Upsilon(1S)\pi^+\pi^-$ and below $0.6GeV$ for the
others. One can see that they are far below the $D$ wave resonant
region ($>1.2GeV$). So as a primary consideration, the $D$ wave
contribution comes only from the $D$ wave terms appeared in
Eq.~(\ref{eq:v0}).

\begin{figure}[htb]
\begin{center}\vspace*{0.5cm}
\begin{center} %\hspace*{1.0cm}
\epsfysize=4cm \epsffile{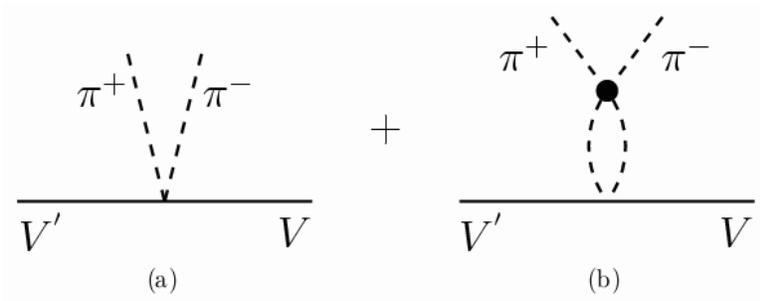}
\caption{\label{fig:feyn1}Diagrams of the vertex
$V^{'}V\pi^+\pi^-$ without $\pi\pi$ FSI (a) and with $\pi\pi$ FSI
(b).}
\end{center}
\end{center}
\end{figure}

The basic diagrams for the $V^{\prime}\to VPP$ decay, where
$V^{\prime}(V)$ and $P$ denote the vector and pseudoscalar mesons,
respectively, are shown in Fig.~\ref{fig:feyn1}. In the figure,
(a) represents the $V^{\prime}\to VPP $ decay without FSI, namely
the tree diagram or Born term, and (b) describes the decay with
$\pi\pi$ FSI. In Fig.~\ref{fig:feyn1}(b), the $\pi\pi\to
\pi^+\pi^-$ $t$-matrix described by the solid black circle is
obtained by the loop resummation \cite{oo97}, namely by a set of
coupled-channel BSEs and both $\pi\pi$ and $K{\bar K}$ channels
are included (for details, see Ref.~\cite{oo97}). To factorize the
$\pi\pi$ FSI from the direct $V'\to VPP$ decay part in
Fig.~\ref{fig:feyn1}(b), the on-shell approximation is adopted.
Thus, only $\pi^+$, $\pi^-$, $\pi^0$ exist in the first loop which
is directly linked to the $V'\to VPP$ vertex. The off-shell
effects can be absorbed into the phenomenological coupling
constants of the vertex. In fact, this approximation is often used
in parameterizing the $S$ wave FSI with phase shift data
\cite{ckk94}. The full $S$ wave $\pi\pi\to \pi^+\pi^-$ $t$-matrix
can be expressed as
%--------------------------------------------------------------------------------
%--------------------------------------------------------------------------------
\begin{equation}  \label{eq:fsi}
\langle \pi^+\pi^- + \pi^-\pi^+
+\pi^0\pi^0|t|\pi^+\pi^-\rangle=2t^{I=0}_{\pi\pi,\pi\pi},
\end{equation}
where $t^{I=0}_{\pi\pi,\pi\pi}$ denotes the full $S$ wave
$\pi\pi\to \pi\pi$ $t$-matrix in the isospin $I=0$ channel, which
is the solution of a set of on-shell coupled channel BSEs
\cite{oo97}. Now, the $t$-matrix for the $V^{'}\to V\pi^+\pi^-$
decay can be expressed as
\begin{equation}
t=V_0 + V_{0S}\cdot G\cdot 2t^{I=0}_{\pi\pi,\pi\pi},
\end{equation}
where $V_0$ is the amplitude of the tree diagram (a), $V_{0S}$ is
the $S$ wave component of $V_0$ and $G$ is the two meson loop
propagator
\begin{equation}
G=i\int\frac{d^4q}{(2\pi)^4}\frac{1}{q^2-m_{\pi}^2+i \varepsilon} \frac{1}{%
(p^{^{\prime}}-p-q)^2-m_{\pi}^2+i\varepsilon}.  \label{eq:2loop}
\end{equation}
The numerical calculation is done by introducing a three-momentum
cutoff $q_{max}$. The value of the cutoff is taken as that used in
\cite{oo97}, where the $\pi\pi$ scattering data can be well
reproduced up to 1.2 GeV, namely, the $\pi\pi$ FSI in our model is
consistent with the $\pi\pi$ scattering data. Moreover, this
cutoff is also consistent with the dimensional regularization
\cite{oop98}. The analytic expression of the loop integral in
Eq.~(\ref{eq:2loop}) can be given as
\begin{equation}
\label{eq:g} G=\frac{1}{8\pi^2}\{\sigma
\arctan{\frac{1}{\lambda\sigma}} -
\ln[\frac{q_{max}}{m_{\pi}}(1+\lambda)]\},
\end{equation}
where $\sigma=\sqrt{\frac{4m_{\pi}^2}{s}-1}$ and
$\lambda=\sqrt{1+\frac{m_{\pi}^2}{q^2_{max}}}$.

The differential decay width with respect to the $\pi\pi$ invariant mass and
$\cos{\theta}^*_\pi$ reads
\begin{equation}
\frac{d\Gamma}{dm_{\pi\pi}dcos{\theta}^*_\pi}=\frac{1}{8M^2(2\pi)^3}
\overline{\sum}\sum|t|^2|{\bf p_1^*}||{\bf p_3}|
\end{equation}
where $\overline{\sum}\sum$ describes the average over initial
states and the sum over final states, and ${\bf p_3}$ is the
3-momentum of the final vector meson in the lab frame.

\section{Results for the $\Psi(2S)\to J/\Psi\pi^+\pi^-$ decay}

\label{sec:psi}

In the model, the parameters involved are the coupling constants
$g_1$, $g_2$ and $g_3$. The values of the parameters can be
determined by fitting the experimental data of the $\Psi(2S)\to
J/\Psi\pi^+\pi^-$ process. It is shown that the resultant $g_3$
value is so small that we can safely take $g_3=0$. The remaining
coupling constants $g_1$ and $g_2$ are obtained by fitting the
total decay rate and the $\pi^+\pi^-$ invariant mass spectrum
simultaneously. The decay data of the $\Psi(2S)\to
J/\Psi\pi^+\pi^-$ process are taken from ref.\cite{bes00}. These
BES data are normalized by using $\Gamma_{\Psi(2S)}= 277keV$ and
$B(\Psi(2S)\to J/\Psi\pi^+\pi^-)=31.7\%$ \cite{pdg04}. The
resultant coupling constants are
\begin{equation}
g_1=0.106, ~~~~~~g_2/g_1=-0.319, ~~~~~~g_3/g_1=0.
\end{equation}
%---------------------------------------------------------------
\begin{figure}[htb]
\begin{center}\vglue 1.cm\hspace*{1.2cm}
\parbox{0.45\textwidth}{\epsfysize=4cm \epsffile{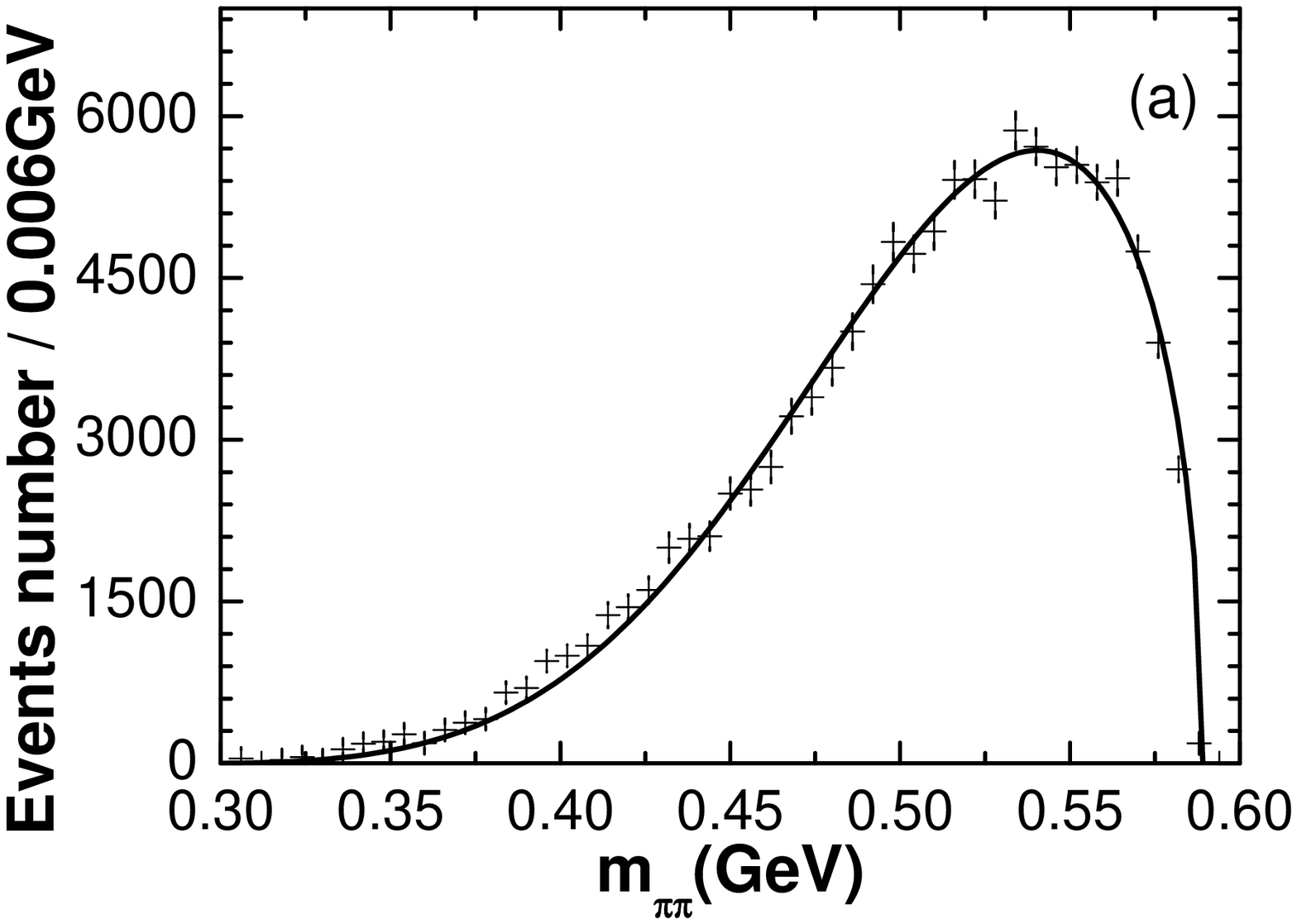}}
\hfill
\parbox{0.45\textwidth}{\epsfysize=4cm \epsffile{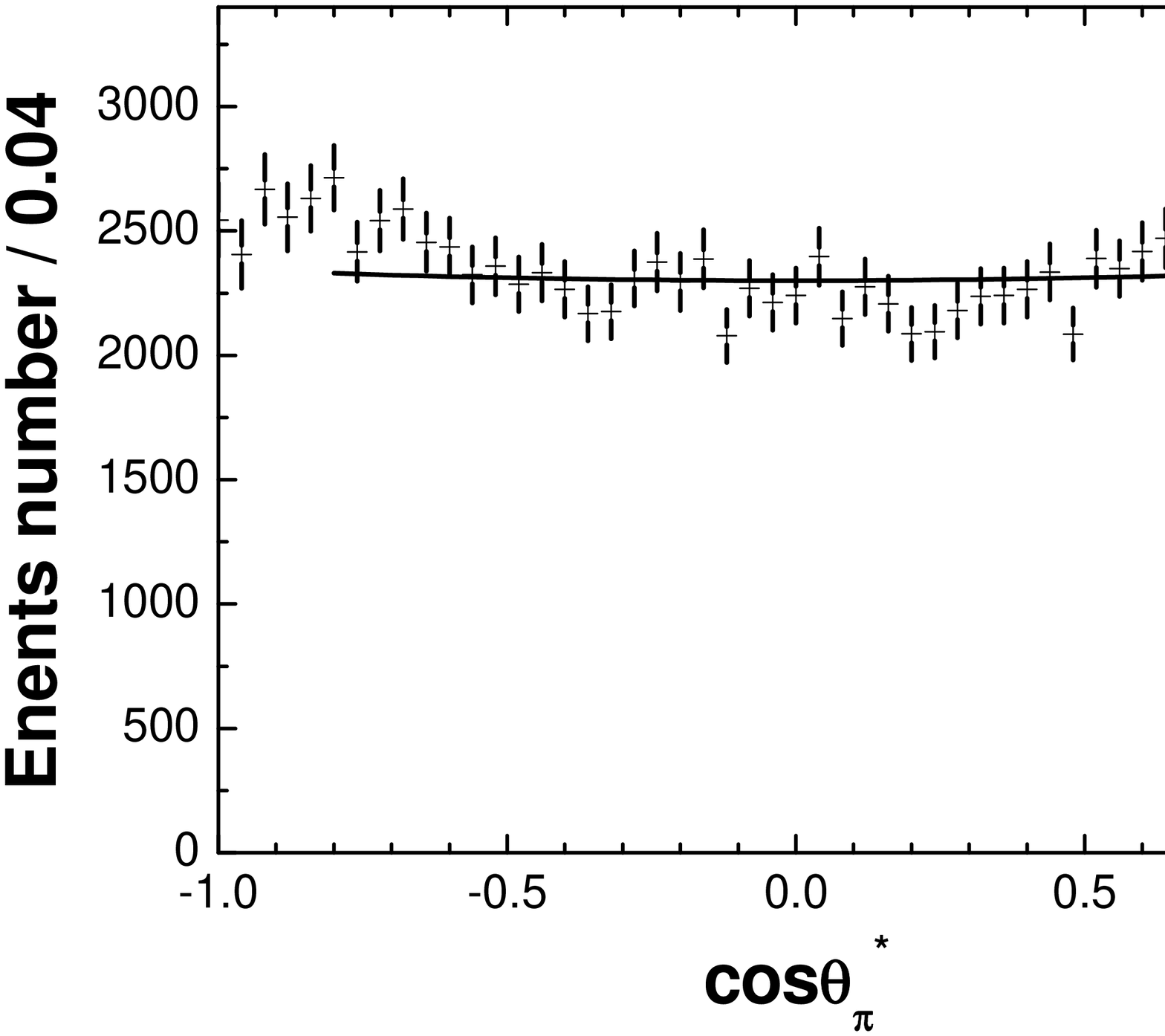}}
\vglue -1.5cm{\caption {\label{fig:psi} $\Psi(2S)\to
J/\Psi\pi^+\pi^-$ decay. (a) $\pi^+\pi^-$ invariant mass spectrum,
and (b) ${\cos \theta_{\pi}^*}$ distribution with $\theta_{\pi}^*$
being the angle between the moving direction of $\pi^+$ in the
$\pi^+\pi^-$ c.m. frame and the moving direction of the
$\pi^+\pi^-$ system in the lab system. The data points are taken
from the BES result \cite{bes00}. The solid curves denote our best
fitted results.}}
\end{center}
\end{figure}
%----------------------------------------------------------------

Our best fit to the $\pi^+\pi^-$ invariant mass spectrum and the
$\cos\theta_{\pi}^*$ distribution are shown in Fig.~\ref{fig:psi}.
It is found that the $\pi^+\pi^-$ invariant mass spectrum is
well-fitted, but the theoretical angular distribution is somewhat
too flat. Note that in fitting angular distribution, we only
consider $\cos\theta_{\pi}^*$ from -0.8 to 0.8, because the
efficiency correction to the data at large $|\cos\theta_{\pi}^*|$ is
not accurate enough \cite{bes00}. The deviation in angular
distribution implies that the $D$ wave contribution is somehow too
small. In fact, as discussed in Section~\ref{sec:for}, in our
calculation, the $D$ wave FSI is not included. It is also found that
the $S$ wave FSI enhances the invariant mass spectrum considerably.
It should be noted that due to $P_2(\cos\theta_{\pi}^*)=
1/2(\cos^2\theta_{\pi}^*-1/3)$, integrating over
$\cos\theta_{\pi}^*$ will result that the $D$ wave contribution is
not so important in the invariant mass spectrum. However, in the
angular distribution, the $\cos\theta_{\pi}^*$-dependence, and
consequently the $D$ wave effect, will explicitly show up. Thus, we
deem that the deviation in angular distribution may be due to lack
of $D$ wave FSI. In fact, it can be confirmed in the following way:
In Ref.\cite{bes00,mu97}, without FSI, the authors can
well-reproduce both the $\pi\pi$ invariant mass spectrum and the
angular distribution with $g_1=0.30{\pm 0.01}$,
$g_2/g_1$$=-0.35\pm0.03$ and $g_3=0$. However, with the $S$ wave
FSI, in the best data fitting to the $\pi\pi$ invariant mass
spectrum, the resultant $g_1$ is 0.106 which is 3 times less than
that in Ref.~\cite{mu97}, while $g_2/g_1$ keeps almost the same
value as that in Ref. \cite{mu97}. This means that the effect of the
$S$ wave FSI is so large that $g_1$ has to be much smaller to
explain the $\pi\pi$ invariant mass spectrum. As a consequence, the
$D$ wave component is also greatly reduced. If we naively multiply a
factor of 3 to the $D$ wave component, the angular distribution can
also be reproduced better. Unfortunately, as mentioned above, the
$D$ wave FSI can not be treated properly in the simple ChUT approach
\cite{oo97}.

\section{The bottomonia $\pi^+\pi^-$ transitions}

\label{sec:upsilon}

Similar to the case of $\Psi(2S)\to J/\Psi\pi^+\pi^-$, in the
$\Upsilon(2S)\to\Upsilon(1S)\pi^+\pi^-$ or
$\Upsilon(3S)\to\Upsilon(2S)\pi^+\pi^-$ decay, $g_3=0$ can be
adopted. But for the $\Upsilon(3S)\to\Upsilon(1S)\pi^+\pi^-$
decay, the $S$ wave FSI is no longer a main contributor, and a
finite value of $g_3$ is requested. With this consideration,
Mannel et al. showed that the resultant values of $g_2/g_1$ for
the $\Upsilon(2S)\to\Upsilon(1S)\pi^+\pi^-$ and
$\Upsilon(3S)\to\Upsilon(2S)\pi^+\pi^-$ processes are very close,
but quite different from that for the
$\Upsilon(3S)\to\Upsilon(1S)\pi^+\pi^-$ process. The latter one is
about ten times larger than the former \cite{mu97}. This is
somewhat unnatural. Suffice to say, the pions involved in
$\Upsilon(3S)\to\Upsilon(1S)\pi^+\pi^-$ are somewhat harder than
in the other bottomonium transitions, and in principle the values
of $g_2/g_1$ for these processes should not be the same for
different dynamical regions. However, in the decay processes
considered, the vector mesons involved are all in the $S$ wave
state, and the particles involved are in the same mass scale, and
the difference among kinematical regions is not too large. Thus we
deem that the values of $g_2/g_1$ for these processes should be
very close. To reduce the number of free parameters, we take the
same $g_2/g_1$ value for different $\Upsilon(nS)$ decays.

%---------------------------------------------------------------------
\begin{figure}[htb]
\begin{center}\vglue 2.cm
\parbox{.33\textwidth}{\epsfysize=4cm \epsffile{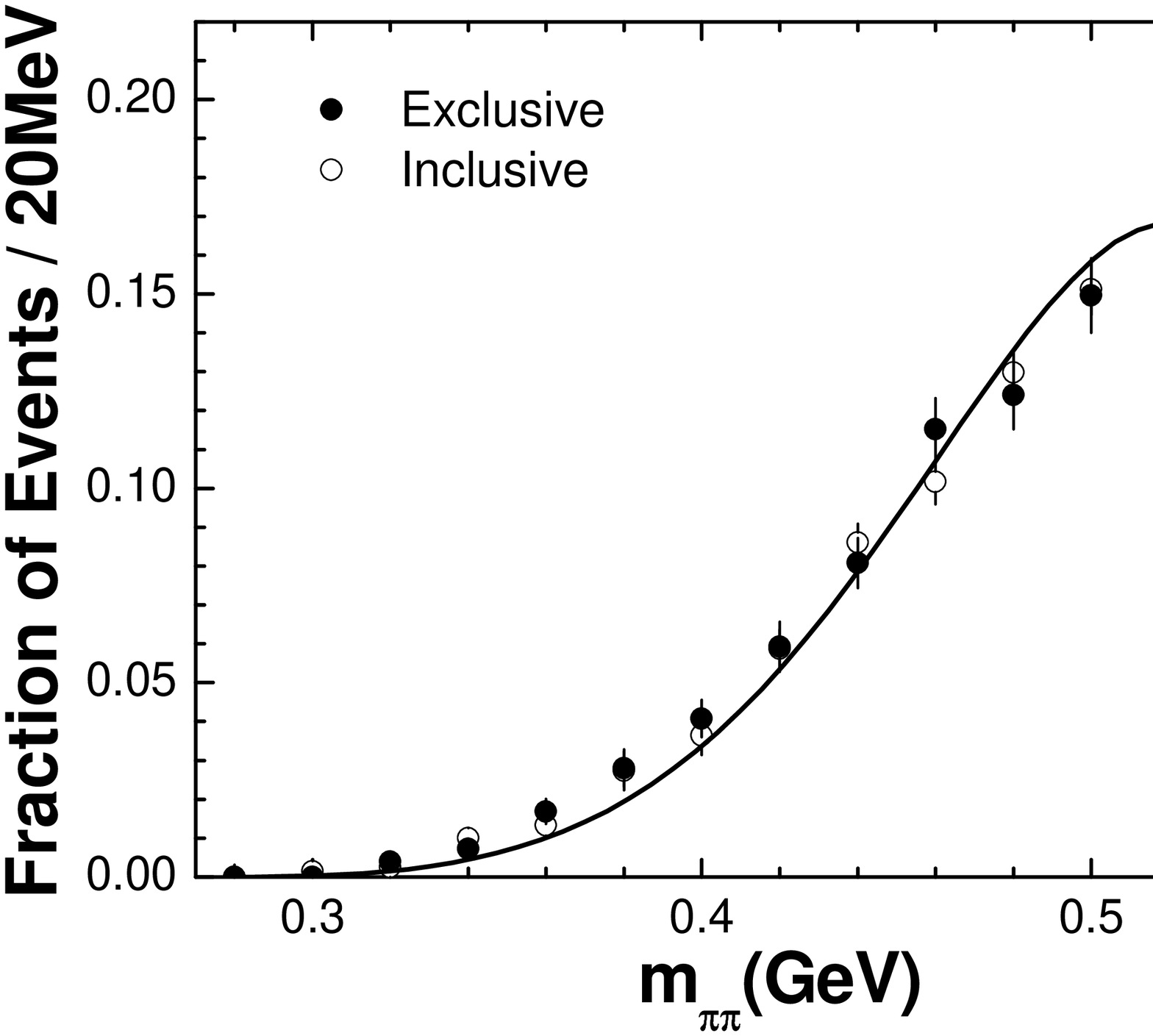}}
\hfill\parbox{.33\textwidth}{\epsfysize=4cm
\epsffile{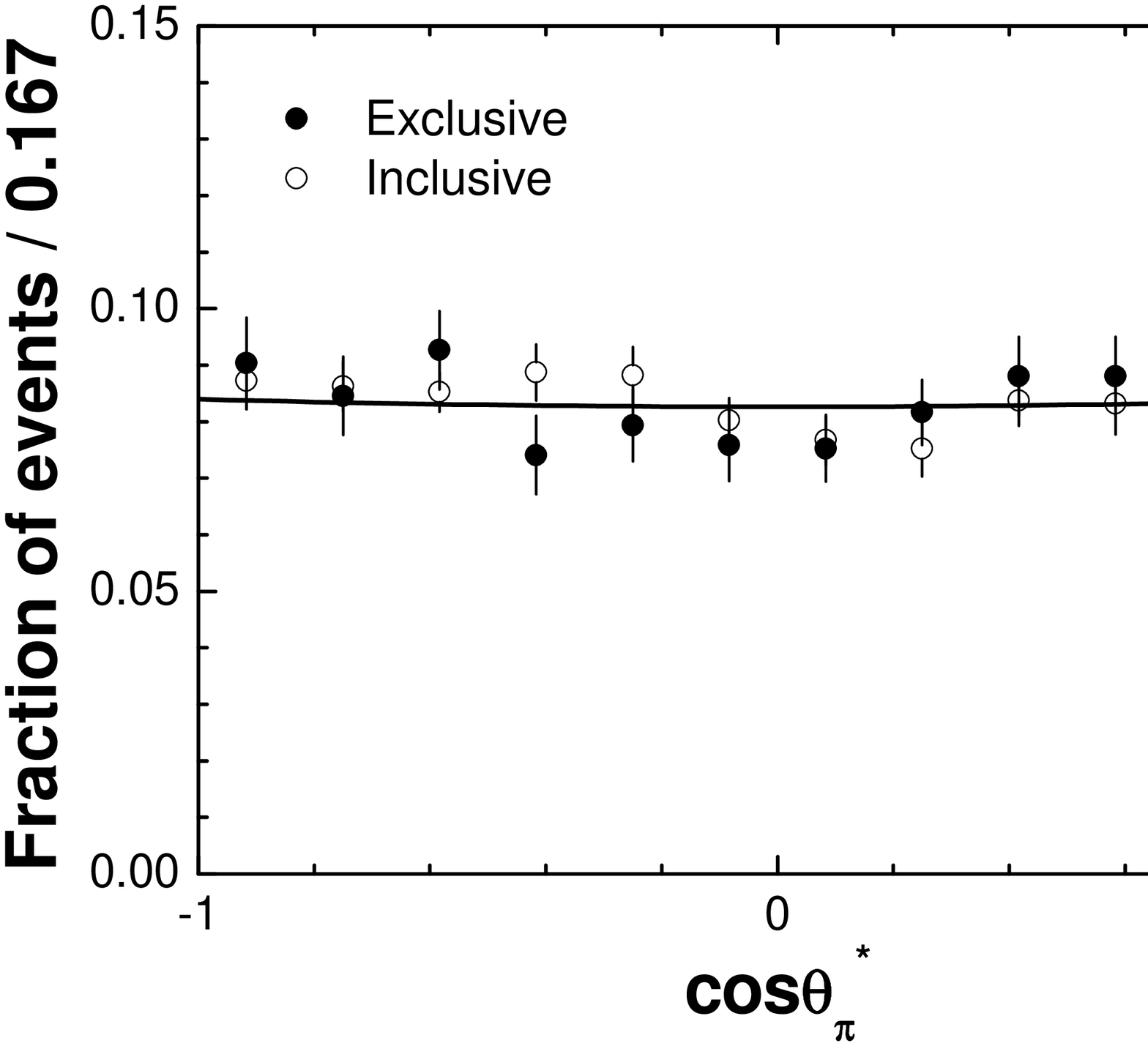}}\hfill
\parbox{.33\textwidth}{\epsfysize=4cm\epsffile{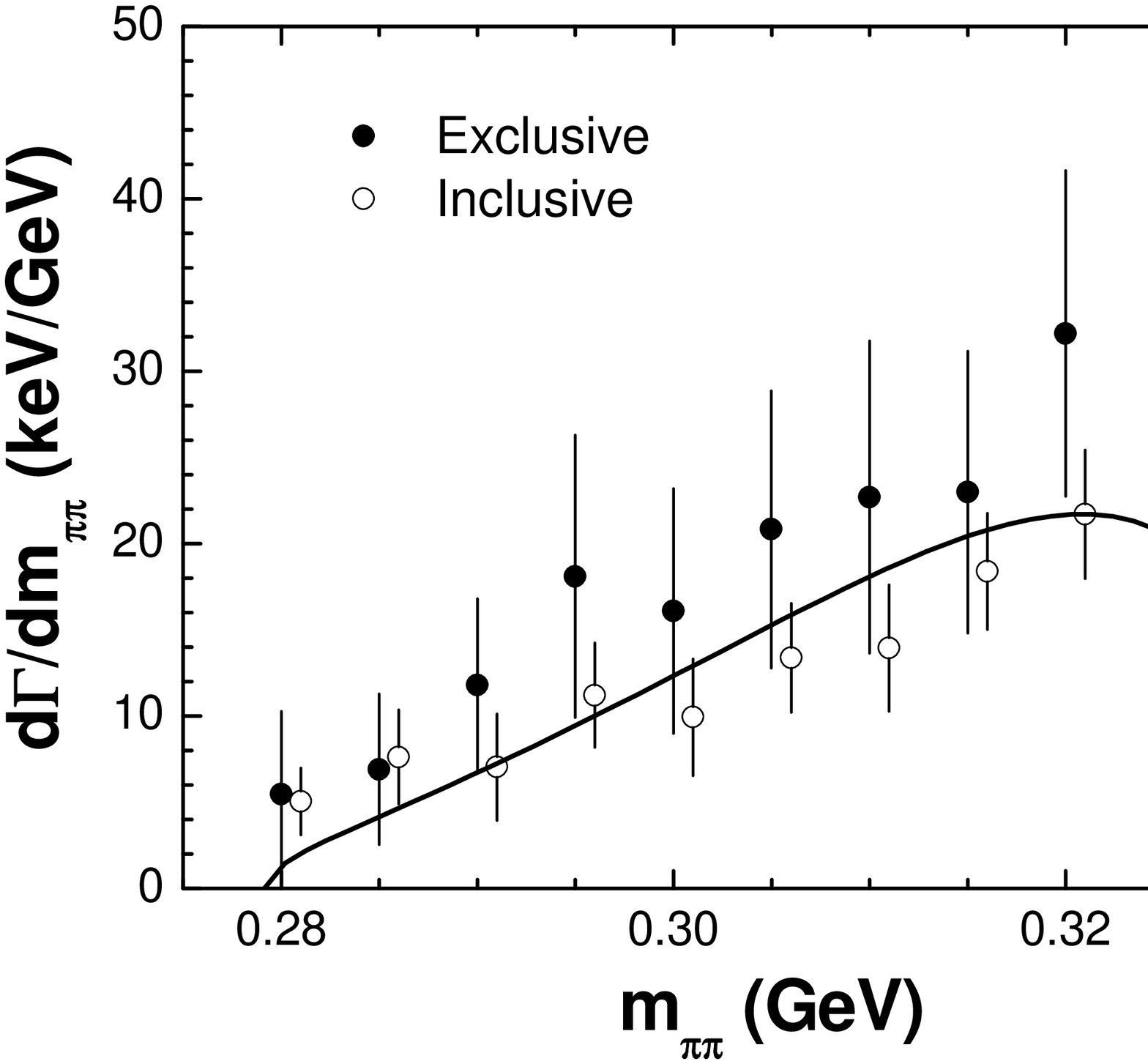}}\hfill
\vglue -1.5cm{\caption{\label{fig:noxfit}(a) The $\pi^+\pi^-$
invariant mass spectrum for the
$\Upsilon(2S)\to\Upsilon(1S)\pi^+\pi^-$ process, (b) the
$\cos\theta_{\pi}^*$ distribution for the
$\Upsilon(2S)\to\Upsilon(1S)\pi^+\pi^-$ process, and (c) the
$\pi^+\pi^-$ invariant mass spectrum for the
$\Upsilon(3S)\to\Upsilon(2S)\pi^+\pi^-$ process.}}
\end{center}
\end{figure}
%---------------------------------------------------------------------
%---------------------------------------------------------------------
\begin{figure}[htb]
\begin{center}\vglue 2.cm \hglue 1.cm
\parbox{.45\textwidth}{\epsfysize=4cm \epsffile{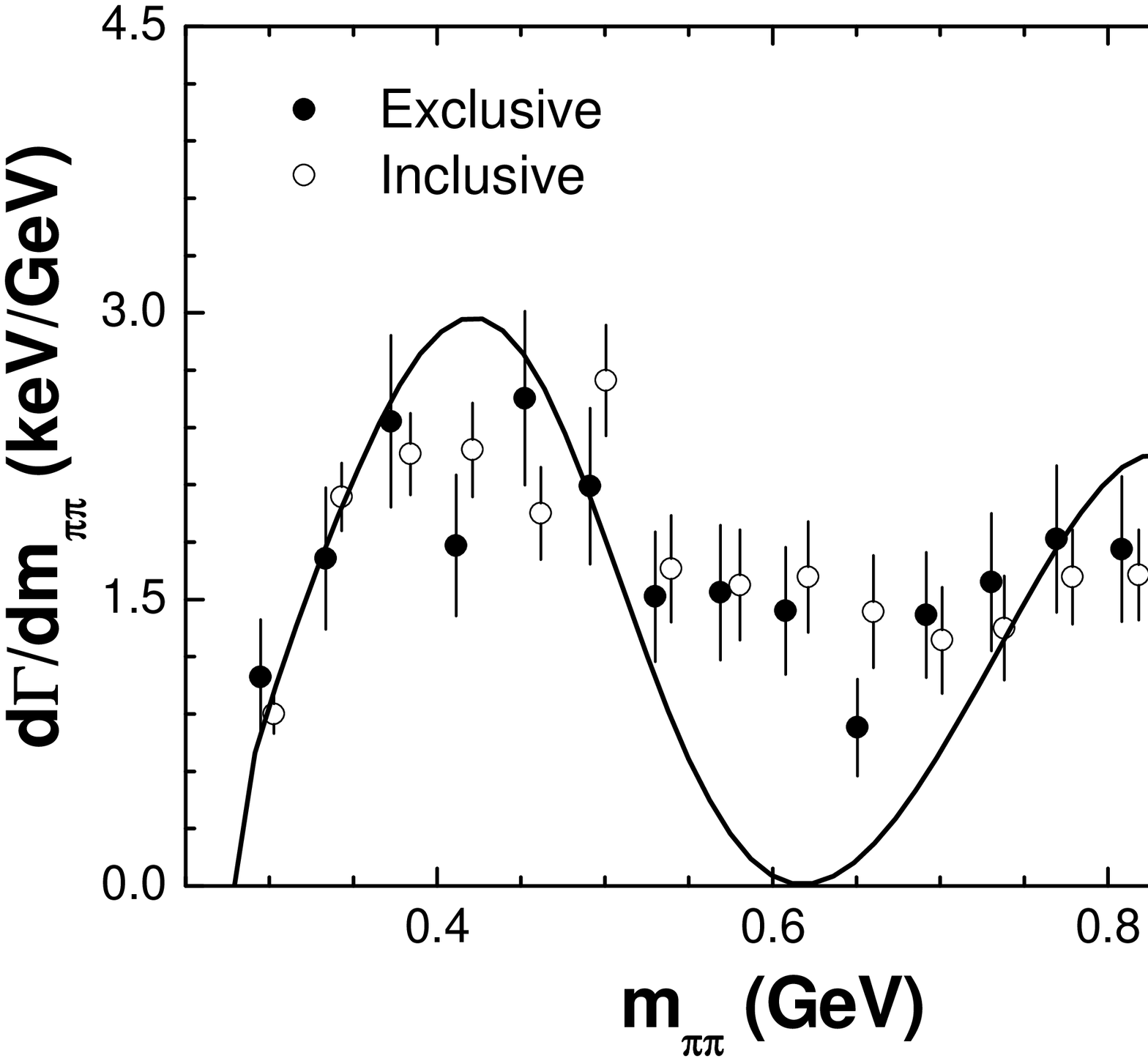}}
\hfill
\parbox{.45\textwidth}{\epsfysize=4cm \epsffile{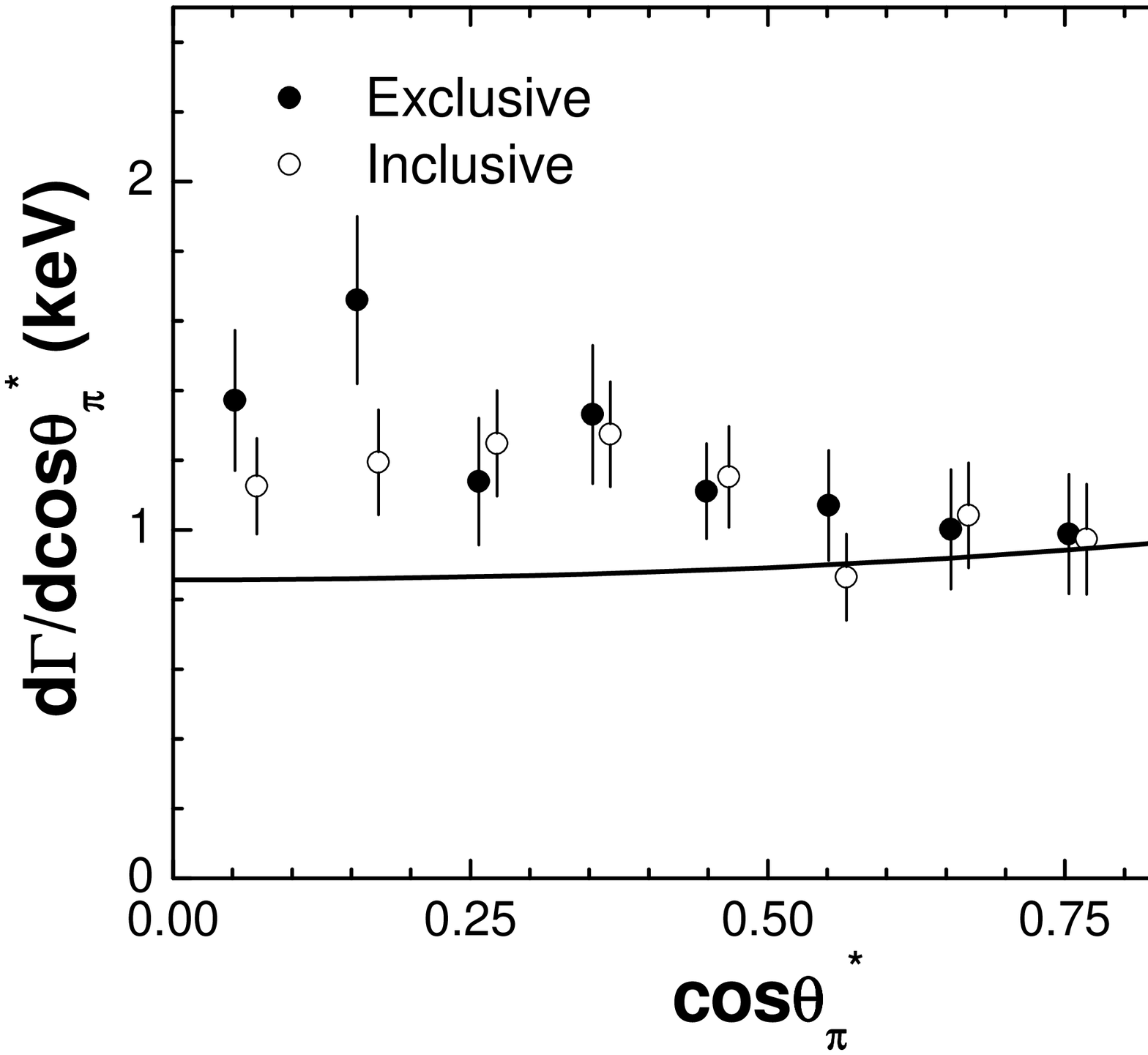}}
\vglue -1.5cm{\caption{\label{fig:nox31fit}(a) The $\pi^+\pi^-$
invariant mass spectrum and (b) the $\cos\theta_{\pi}^*$
distribution for the $\Upsilon(3S)\to\Upsilon(1S)\pi^+\pi^-$
process.}}
\end{center}
\end{figure}
%---------------------------------------------------------------------

The decay data for $\Upsilon(2S)\to\Upsilon(1S)\pi^+\pi^-$ are
taken from \cite{cleo98} and for
$\Upsilon(3S)\to\Upsilon(2S)\pi^+\pi^-$ and
$\Upsilon(3S)\to\Upsilon(1S)\pi^+\pi^-$ from \cite{cleo94}. To get
the physical coupling constants, the data for
$\Upsilon(2S)\to\Upsilon(1S)\pi^+\pi^-$ \cite{cleo98} is
normalized by $\Gamma_{\Upsilon(2S)}= 43keV$ and
$B(\Upsilon(2S)\to\Upsilon(1S)\pi^+\pi^-)=19.2\%$ \cite{pdg04}.
Our calculated results are plotted in
Figs.~\ref{fig:noxfit}-\ref{fig:nox31fit}. It is shown that the
resultant $\pi\pi$ invariant mass spectra for both
$\Upsilon(2S)\to\Upsilon(1S)\pi^+\pi^-$ and
$\Upsilon(3S)\to\Upsilon(2S)\pi^+\pi^-$ decays agree with the data
values, but the angular distribution for the former one is
somewhat flat, which might also be due to the same reason
discussed in Section~\ref{sec:psi}. On the other hand, there is
almost no way to fit the $\pi^+\pi^-$ invariant mass spectrum and
the $\cos\theta_{\pi}^*$ distribution of the
$\Upsilon(3S)\to\Upsilon(1S)\pi^+\pi^- $ process simultaneously,
even if $g_2/g_1$ is further released as a free parameter. The
resultant parameters are listed in Tab.~\ref{tab:nox}.
\begin{table}[ht]
\begin{center}
\caption{ Resultant parameters through data fitting.}
\begin{tabular}{|c|c|c|c|}\hline
Decay &$g_1$ &$g_2/g_1$ & $g_3/g_1$ \\ \hline
$\Upsilon(2S)\to\Upsilon(1S)\pi^+\pi^-$  & 0.0944 & -0.230 & 0 \\
\hline
 $\Upsilon(3S)\to\Upsilon(2S)\pi^+\pi^-$  & 0.768  & -0.230 & 0 \\ \hline
 $\Upsilon(3S)\to\Upsilon(1S)\pi^+\pi^-$  & 0.0123 & 0.564 & -13.602
 \\ \hline
\end{tabular}
\label{tab:nox}
\end{center}
\end{table}

\subsection{Sequential decay mechanism}

\begin{figure}[htb]
\begin{center}\vspace*{0.5cm}
\begin{center} %\hspace*{1.0cm}
\epsfysize=4.cm \epsffile{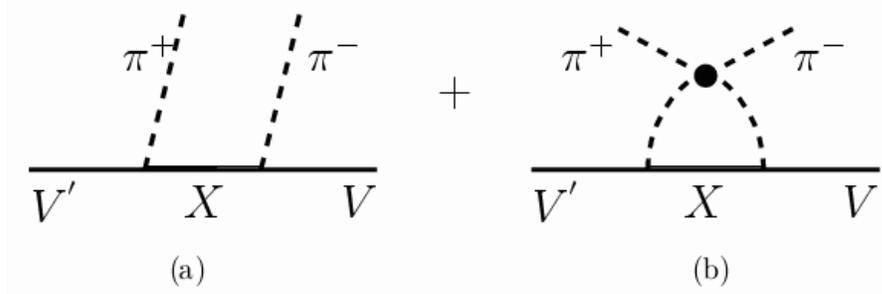} {\caption{\label{fig:feyn2}
Diagrams of the sequential $X$ decay mechanism (a) at tree level,
and (b) with the $\pi\pi$ FSI.}}
\end{center}
\end{center}
\end{figure}

In order to explain the decay data of the
$\Upsilon(3S)\to\Upsilon(1S)\pi^+\pi^-$ decay, we propose an
additional sequential decay mechanism where an intermediate state,
called $X$, is introduced. Additional Feynman diagrams for
$\Upsilon(nS)\to\Upsilon(mS)\pi^+\pi^-$ are shown in
Fig.~\ref{fig:feyn2}, where (a) depicts the tree diagram and (b)
the diagram including the $\pi\pi$ $S$ wave FSI.
%--------------------------------------------------------------------------------
%--------------------------------------------------------------------------------
We adopt a simple $S$ wave coupling for $\Upsilon(nS)\to\pi X$.
The quantum numbers of $X$ should be $J^P=1^+$ and $I=1$. The
decay amplitude of Fig.~\ref{fig:feyn2}(a) can be written as
\begin{equation}
V_X^{tree}=g_{nm} \epsilon^{^{\prime}}_{\mu}\epsilon^{*}_{\nu}(\frac{%
-g^{\mu\nu}+p_{X^+}^{\mu}p_{X^+}^{\nu}/m_X^2}{p_{X^+}^2-m_X^2+i m_X\Gamma_X}+%
\frac{-g^{\mu\nu}+p_{X^-}^{\mu}p_{X^-}^{\nu}/m_X^2}{p_{X^-}^2-m_X^2+i
m_X\Gamma_X})  \label{eq:xtree}
\end{equation}
where $p_{X^+}$ and $p_{X^-}$ are the momenta of $X^+$ and $X^-$
respectively, $g_{nm}$ is an effective coupling constant among
$\Upsilon(nS)$, $\Upsilon(mS)$, $\pi^+$ and $\pi^-$ via an
intermediate resonant state $X$. In fact, $g_{nm}$ is the product
of two coupling constants $g_{nX}$ and $g_{mX}$ where
$g_{kX}(k=n,m)$ denotes the coupling constant for the
$\Upsilon(kS)X\pi$ vertex. To further consider the effect of the
$\pi\pi$ $S$ wave FSI, the contribution of Fig.~\ref{fig:feyn2}(b)
should be included. In this figure, the three-propagator loop can
be expressed as
\begin{equation}
G_X^{\mu\nu}=i\int\frac{dq^4}{(2\pi)^4} \frac{-g^{\mu\nu}+p_{X}^{\mu}p_{X}^{%
\nu}/m_X^2}{p_X^2-m_X^2+i\varepsilon} \frac{1}{q^2-m_{\pi}^2+i \varepsilon}
\frac{1}{(p^{^{\prime}}-p-q)^2-m_{\pi}^2+i\varepsilon}  \label{eq:3loop}
\end{equation}
where $p_X=p^{^{\prime}}-q$ is the four-momentum of $X$. The
calculation is carried out in the c.m. frame of the $\pi\pi$
system with the same cutoff value used in the two-meson loop
calculation. As argued in ref.\cite{absz95}, terms with
$\epsilon^{^{\prime}}_{\mu}\epsilon^{*}_{\nu}p_X^{\mu}p_X^{\nu}/m_X^2$
in Eqs.~\ref{eq:xtree} and \ref{eq:3loop} can be neglected,
because of the expected heavy mass of $X$. Then the total
$t$-matrix can finally be written as
\begin{equation}  \label{eq:total}
t=V_0 + V_{0S}\cdot G\cdot 2t^{I=0}_{\pi\pi,\pi\pi} + V_X^{tree} +
g_{nm}\epsilon^{^{\prime}}_{\mu}\epsilon^{*}_{\nu} G_X^{\mu\nu}\cdot
2t^{I=0}_{\pi\pi,\pi\pi}.
\end{equation}

\subsection{Results for bottomonium $\pi^+\pi^-$ transitions}

\label{subsec:Xre} In terms of the $t$-matrix in
Eq.~(\ref{eq:total}), we calculate the $\pi^+\pi^-$ invariant mass
spectra and the $\cos\theta_{\pi}^*$ distributions of the
bottomonium $\pi^{+}\pi^{-}$ transitions. Similar to the argument
given in the former sections, we take $g_3=0$ for the
$\Upsilon(2S)\to\Upsilon(1S)\pi^+\pi^-$ and
$\Upsilon(3S)\to\Upsilon(2S)\pi^+\pi^-$ decays and $g_3$ as a free
parameter for the $\Upsilon(3S)\to\Upsilon(1S)\pi^+\pi^-$ process.
We also demand the values of $g_2/g_1$ to be the same for all
three decays for reducing the number of free parameters. The
values of $g_1$ and $g_2$ are determined by fitting the
experimental decay data \cite{cleo98,cleo94}.

%----------------------------------------------------------
\begin{figure}[htb]
\begin{center}\vglue 2.cm
\parbox{.33\textwidth}{\epsfysize=4cm \epsffile{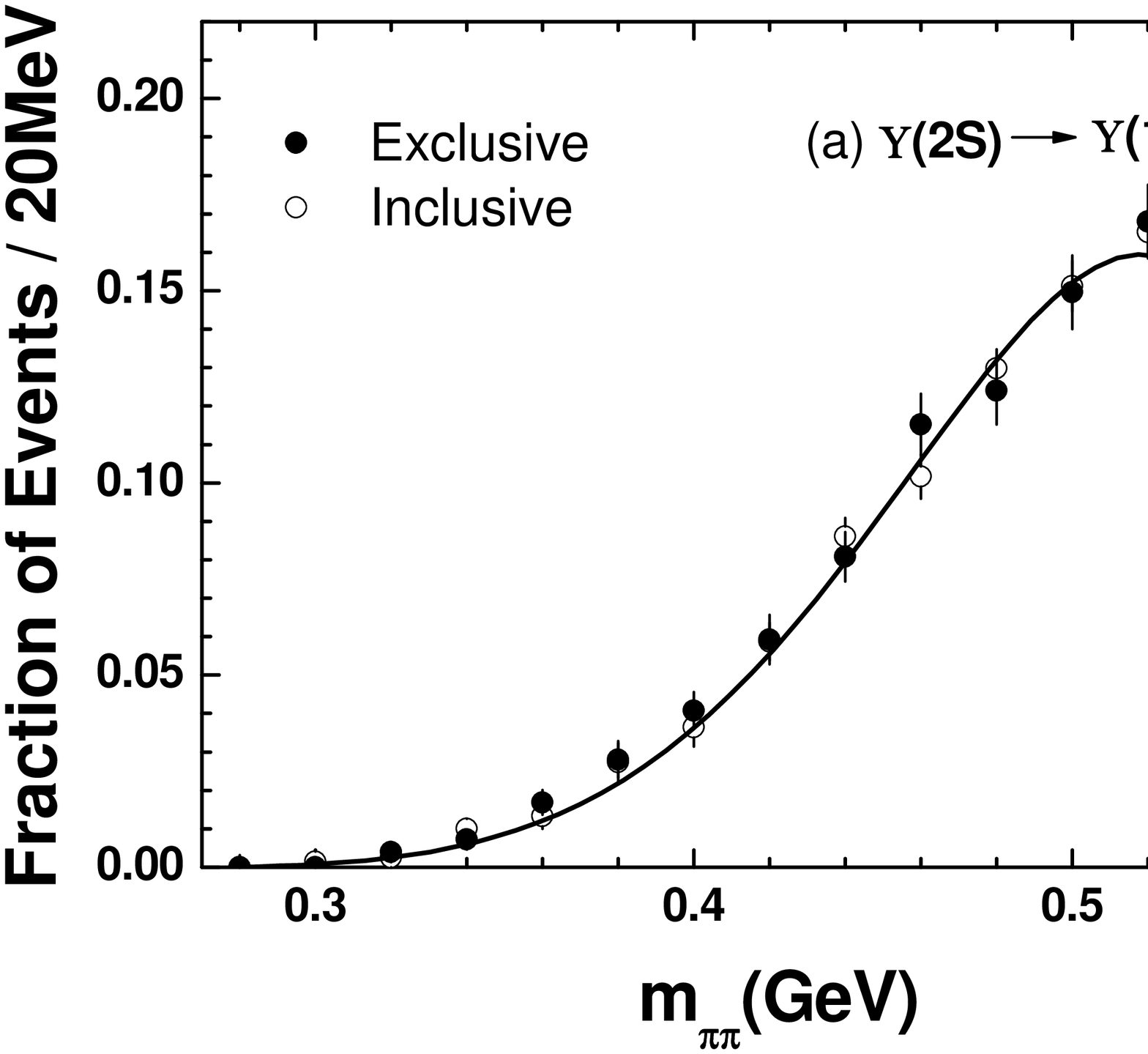}}
\hfill
\parbox{.33\textwidth}{\epsfysize=4cm \epsffile{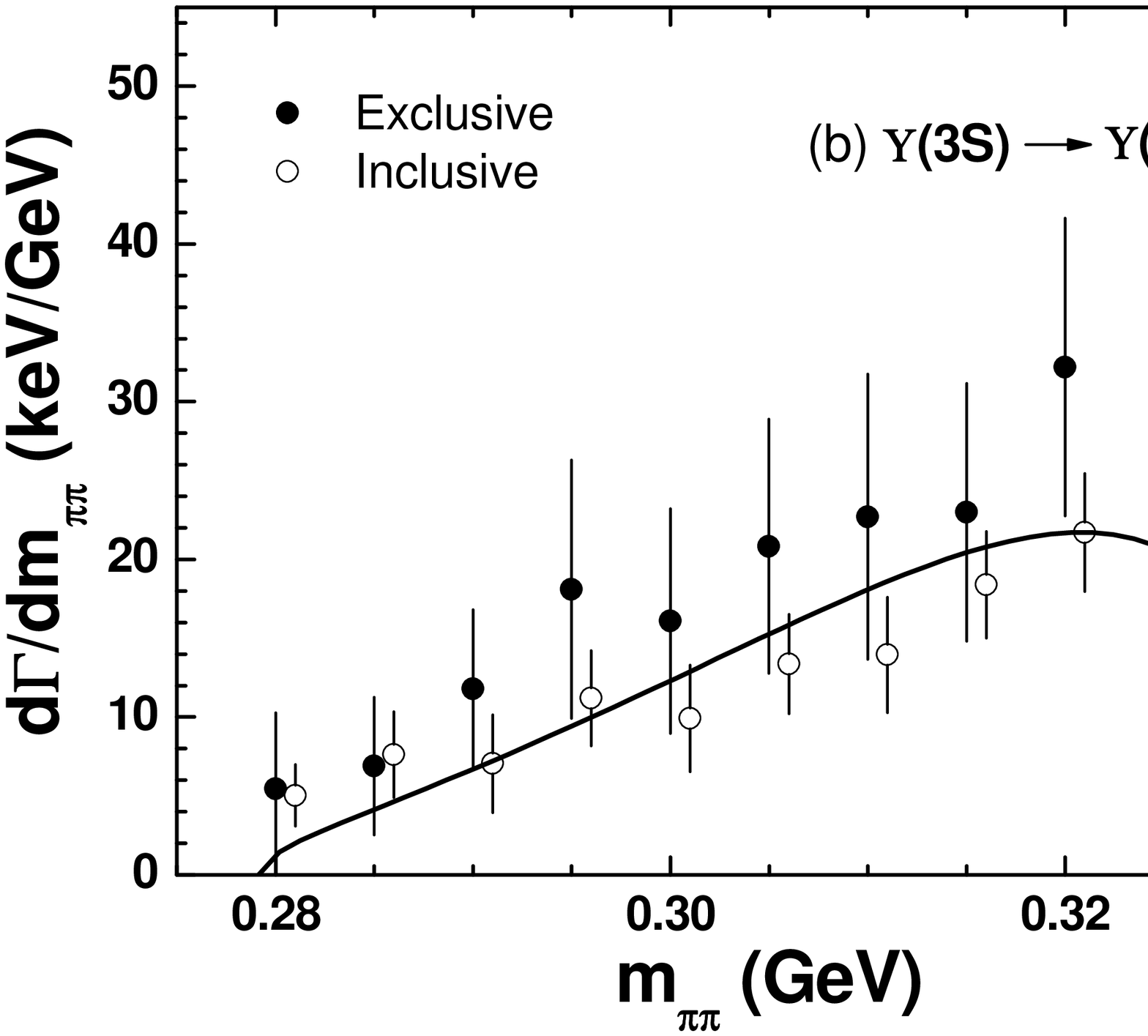}}\hfill
\parbox{.33\textwidth}{\epsfysize=4cm\epsffile{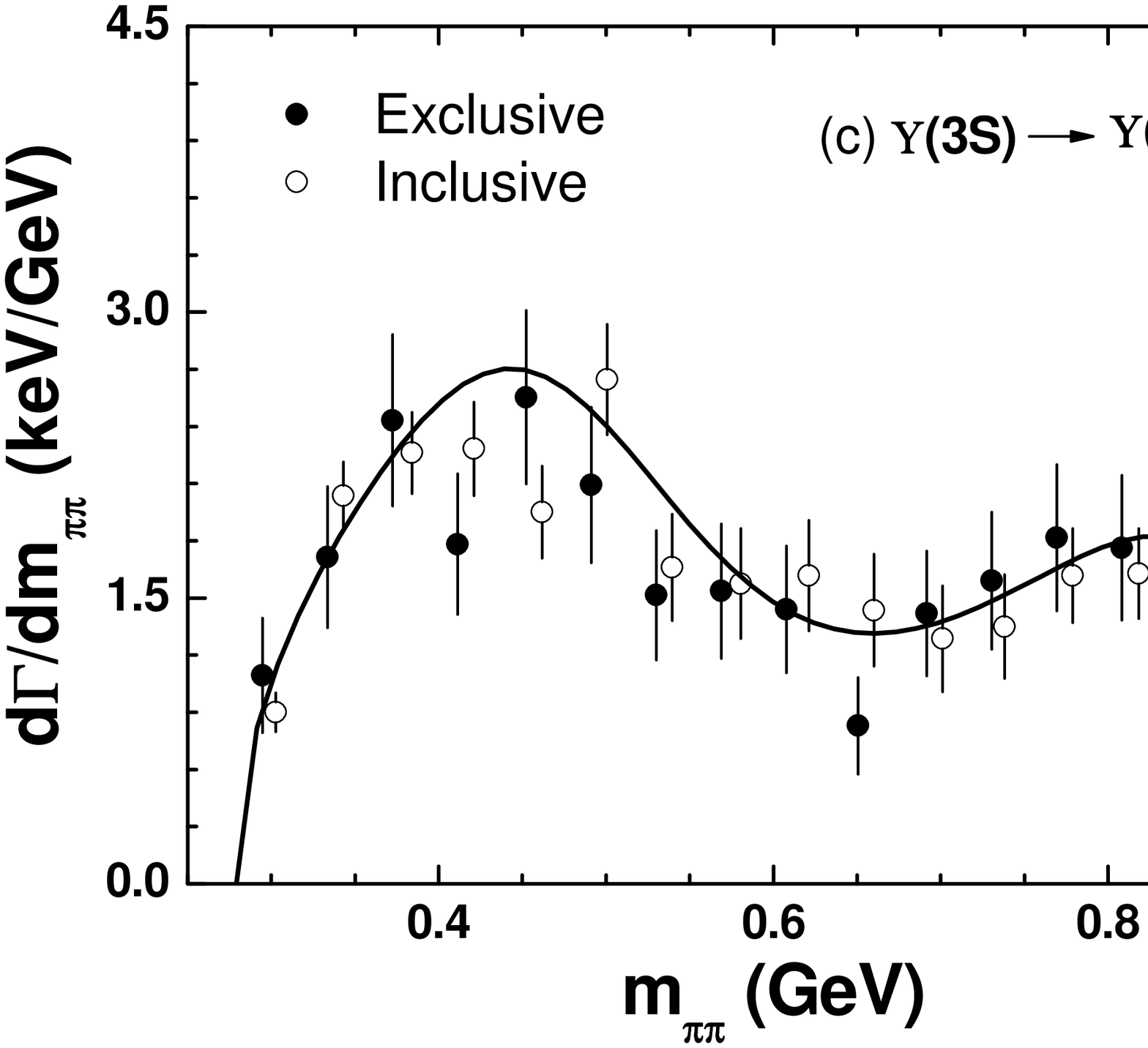}}
\hfill
\parbox{.33\textwidth}{\epsfysize=4cm \epsffile{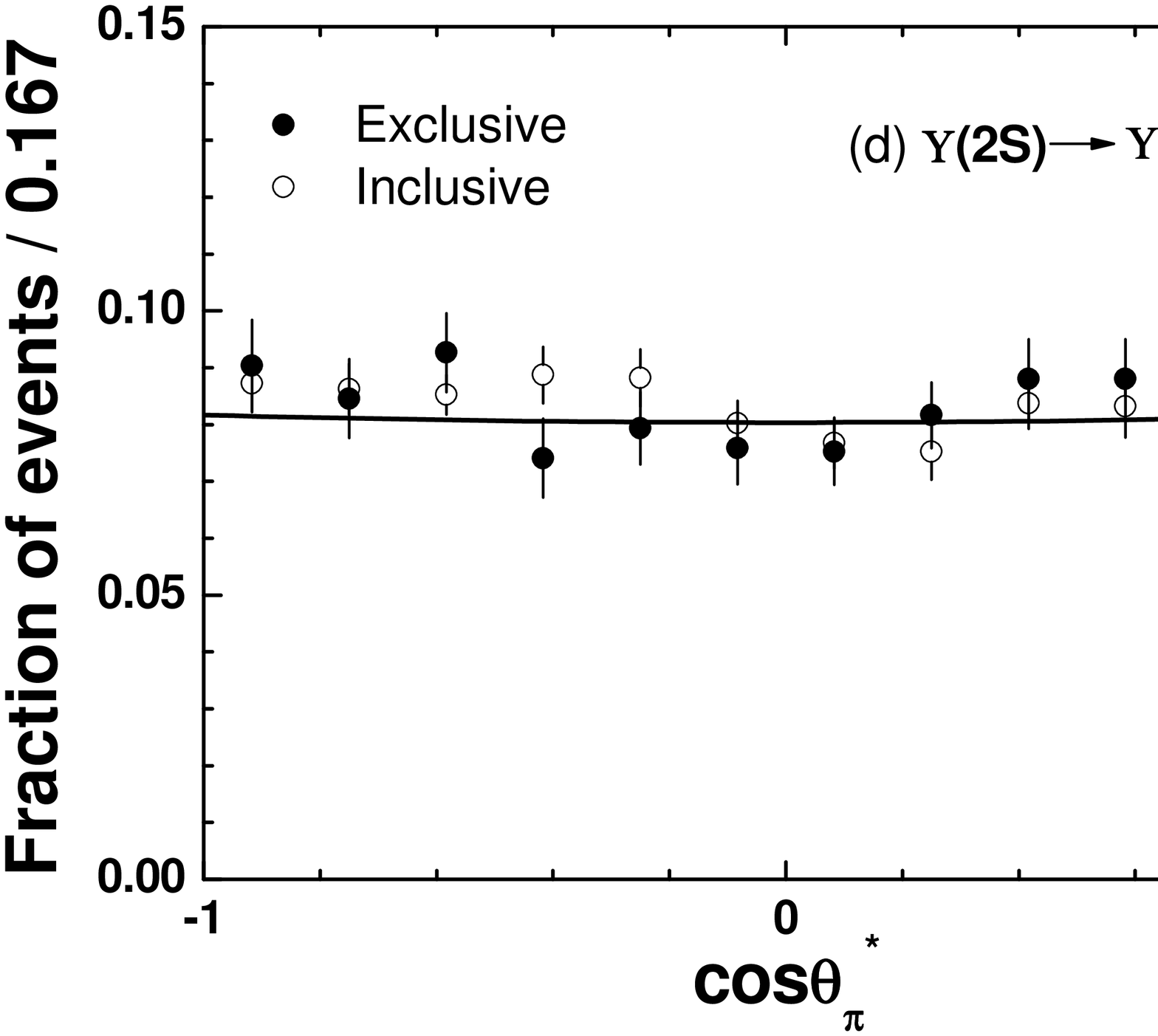}}\hfill
\parbox{.33\textwidth}{\epsfysize=4cm\epsffile{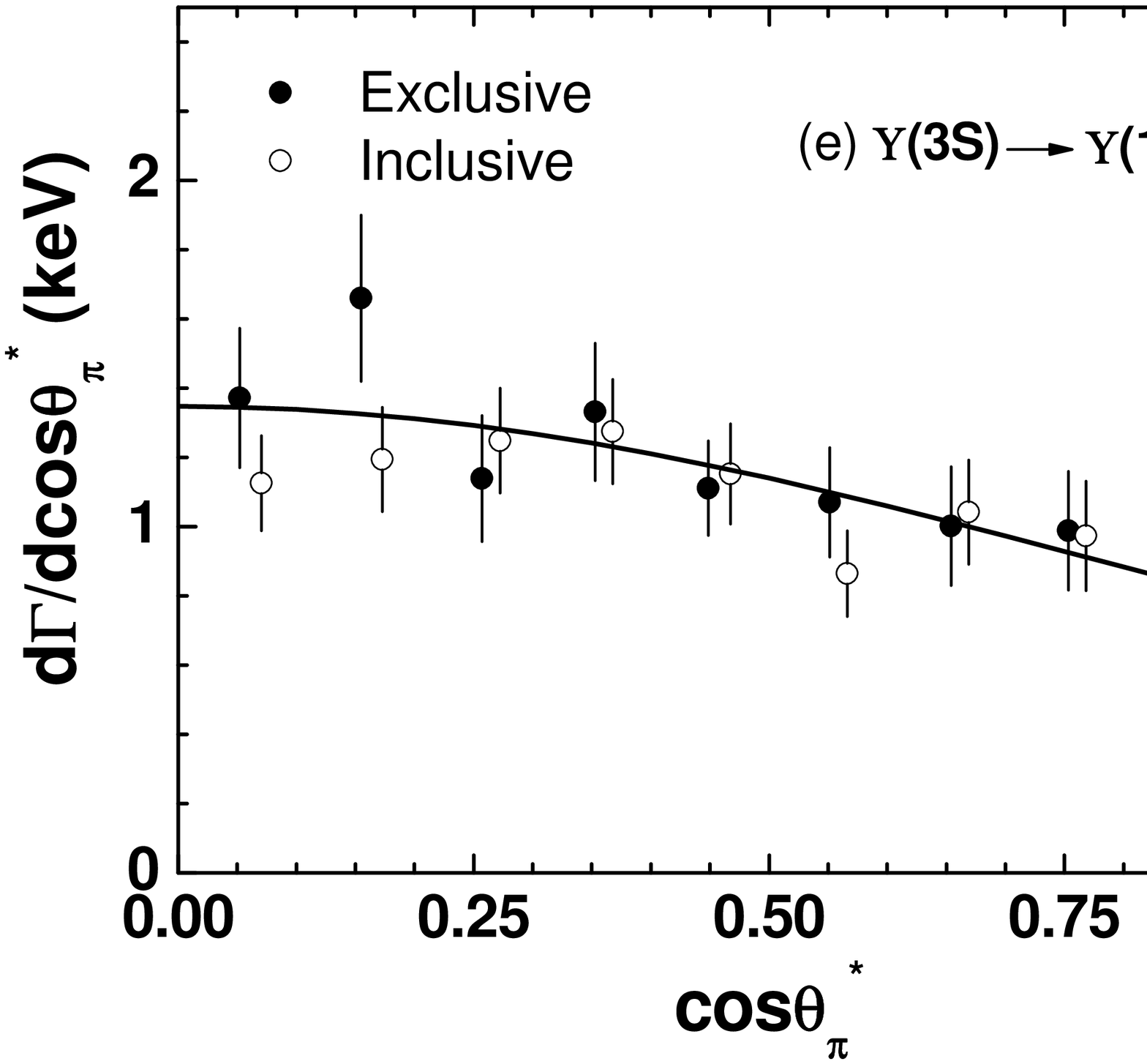}}
\vglue -1.5cm{\caption{\label{fig:fit} The $\pi\pi$ invariant mass
spectra and the $\cos\theta_{\pi}^*$ distributions in the
$\Upsilon(nS)\to \Upsilon(mS)\pi^+\pi^-$ decays.}}
\end{center}
\end{figure}
%----------------------------------------------------------
The calculated results are plotted in Fig.~\ref{fig:fit}. It is
shown that not only both the $\pi\pi$ invariant mass spectrum and
the $\cos\theta_\pi^*$ distribution of the
$\Upsilon(3S)\to\Upsilon(1S)\pi^+\pi^-$ process can simultaneously
be well explained, but also a consistent description of other
bottomonium $\pi^+\pi^-$ transitions can be obtained. The
resultant parameters are tabulated in Tab.~\ref{tab:fit}.
\begin{table}[hbt]
\begin{center}
\caption{\label{tab:fit} Resultant parameters in the data fitting.
In the first column, $n\to m$ denotes the
$\Upsilon(nS)\to\Upsilon(mS)\pi^+\pi^-$ decay. }
\begin{tabular}{|c|c|c|c|c|c|c|}
\hline Decay & $g_1$ & $g_2/g_1$ & $g_3/g_1$ & $g_{nm}$(GeV$^2$) &
$m_X$(GeV) & $\Gamma_X$(GeV)  \\ \hline $2\to1$ & 0.0886 & -0.230 &
0 & -2.316 &  &    \\ \cline{1-5}
$3\to 2$ & 0.769 & -0.230 & 0 & -0.00418 & 10.080 & 0.655   \\
\cline{1-5} $3\to 1$ & 0.00546 & -0.230 & 4.949 & 4.712 &  &   \\
\hline
\end{tabular}
\end{center}
\end{table}
%----------------------------------------------------------

%------------------------------------------------------------
\begin{figure}[htb]
\begin{center}\vglue 2.cm\hspace*{1.0cm}
\parbox{.45\textwidth}{\epsfysize=4cm \epsffile{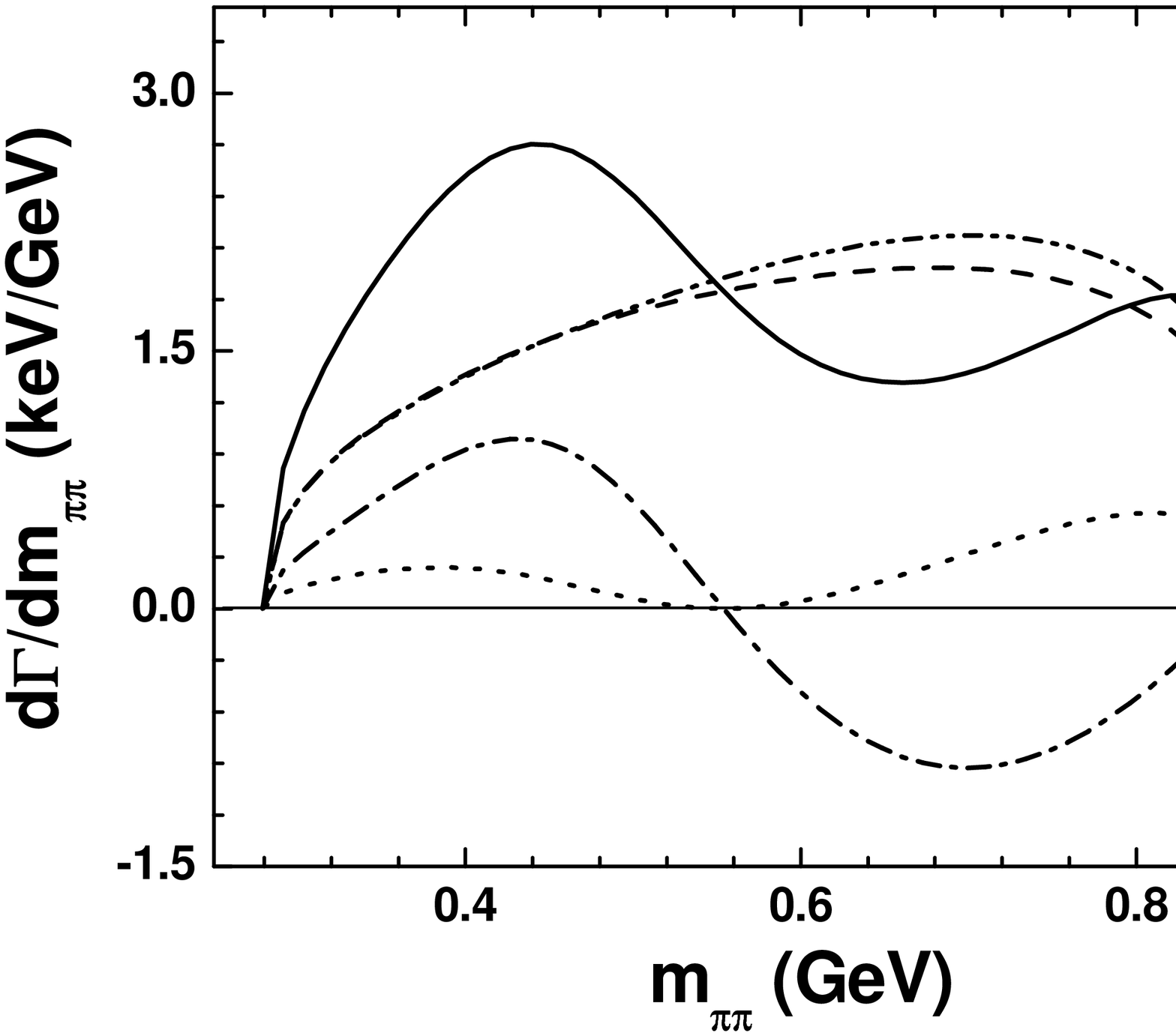}}
\hfill
\parbox{.45\textwidth}{\epsfysize=4cm \epsffile{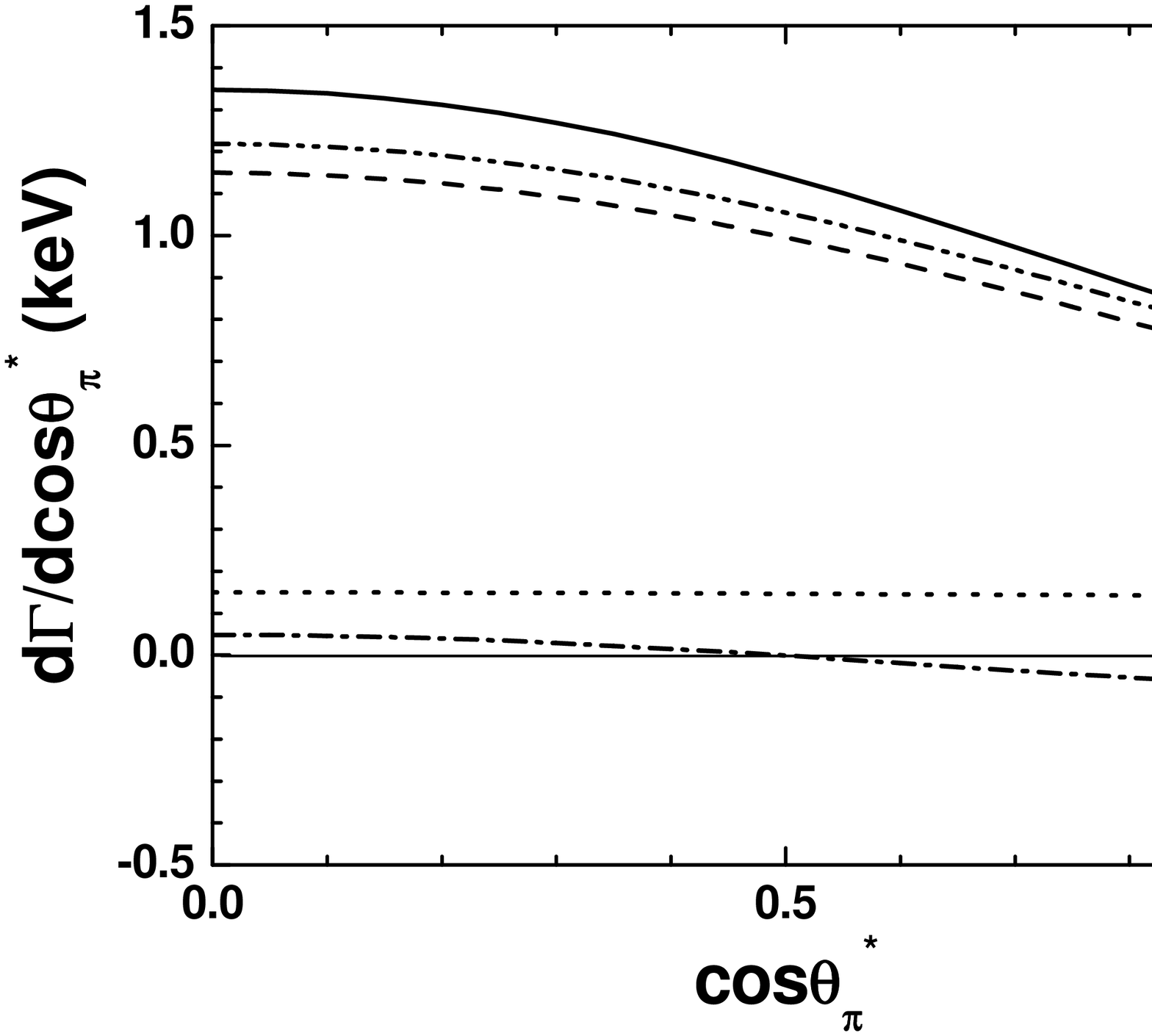}}
\vglue -1.5cm{\caption {\label{fig:analyze} The contributions from
different components to (a) the $\pi^+\pi^-$ invariant mass
spectrum and (b) the $\cos\theta_\pi^*$ distribution in the
$\Upsilon(3S)\to\Upsilon(1S)\pi^+\pi^-$ decay. The solid curves
represent our best fitted results, the dotted, dashed and
dash-dotted curves describe the contributions from terms without
$X$, with $X$ only and the interference term, respectively. The
dash-dot-dotted curves represent the tree level contributions with
$X$ only.}}
\end{center}
\end{figure}
%------------------------------------------------------------

To understand thoroughly the roles of different terms in the
$\pi^{+}\pi^{-}$ invariant mass spectrum and the
$\cos\theta_{\pi}^{\ast}$ distribution of the
$\Upsilon(3S)\to\Upsilon(1S)\pi^{+}\pi^{-}$ decay, it is necessary
to analyze their individual contributions. The results are shown
in Fig.~\ref{fig:analyze}. In the figure, the solid curves
represent our best fitted results, and the dotted, dashed, and
dash-dotted curves describe the contributions from the terms
without $X$ and with $X$ only and the interference term
respectively, and the dash-dot-dotted curves represent the tree
level contributions with $X$ only. The calculated $\pi\pi$
invariant mass spectrum (Fig.~\ref{fig:analyze} (a)) shows that
the contribution from $X$ plays a dominant role, the contribution
from the terms without $X$ can qualitatively but not
quantitatively give the two-peak feature, and the interference
term contributes constructively in the smaller $\pi\pi$ invariant
mass region but destructively in the larger invariant mass region.
The resultant $\cos\theta_{\pi}^{\ast}$ distribution
(Fig.~\ref{fig:analyze} (b)) further shows that the contribution
from $X$, even in the tree level, produces almost the whole
angular distribution structure. Although the scalar meson $\sigma$
dynamically generated by the $S$ wave $\pi\pi$ FSI in ChUT
\cite{oo97} can make a peak around its pole position at about 450
MeV in the $\pi\pi$ invariant mass spectrum, the contribution from
the diagrams without $X$ is not dominant due to smaller values of
coupling constants. Thus, an additional $D$ wave FSI which
provides a flat contribution in the invariant mass region
considered will not be an important contributor. These indicate
that the intermediate state $X$ is very important in reproducing
not only the $\pi^{+}\pi^{-}$ invariant mass spectrum but also the
$\cos\theta_{\pi }^{\ast}$ distribution in the
$\Upsilon(3S)\to\Upsilon(1S)\pi^{+}\pi^{-}$ decay.

If we further consider the quark structures of the particles
involved, the intermediate state should contain $b,~\bar{b},~q$
and $\bar{q}$. This state might be a tetraquark state, for
instance, $b\bar{b}u\bar{d}$, with $J^P=1^+$ and $I=1$ for $X^+$
or a $B\bar{B}$ bound state, for instance, $B^+\bar{B}^0$, for
$X^+$.

%--------------------------------------------------------------------------------
\begin{figure}[htb]
\begin{center}\vspace*{0.5cm}
\begin{center} %\hspace*{1.0cm}
\epsfysize=6.cm \epsffile{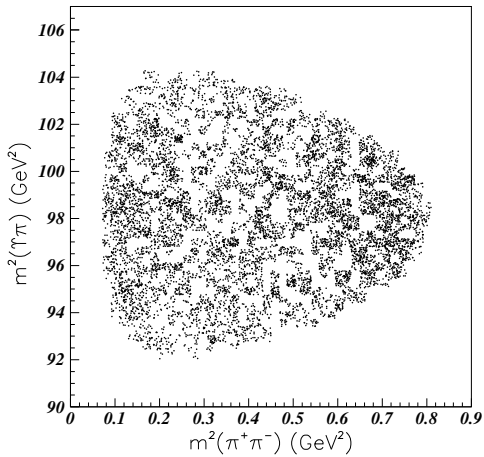} {\caption{\label{fig:dalitz}
Dalitz plot for $\Upsilon(3s)\to\Upsilon(1s)\pi^+\pi^-$. The mass
squared of $\Upsilon(1S)\pi$ system is plotted against the mass
squared of the $\pi^+\pi^-$ system.}}
\end{center}
\end{center}
\end{figure}
%--------------------------------------------------------------------------------

It should be mentioned that similar mechanism was also proposed by
V.V. Anisovich {\it et al.}\cite{absz95}. In their paper, a
trivial $S$ wave coupling was used in the effective vertex
$\Upsilon(nS)\Upsilon(mS)\pi\pi$ which is described by
Eq.~(\ref{eq:v0}) in our model with both $S$ wave and $D$ wave
components. With that mechanism, they successfully reproduced the
$\pi\pi$ invariant mass spectra of the
$\Upsilon(nS)\to\Upsilon(mS)\pi^+\pi^-$ decays, but did not give
reasonable $\cos\theta_{\pi}^*$ distributions due to the dominance
of the $\pi\pi$ $S$ wave in their model. As a result, the
estimated mass of the additional intermediate state is in the
range of 10.4-10.8 GeV which is located outside the data area of
the Dalitz plot where the data points in the direction of
$m_{\Upsilon\pi}$ are located from 9.6 GeV to 10.2 GeV. Thus the
effect of the state does not show up in the Dalitz plot.

The mass and width of the intermediate state in our work are
different from those in Ref.~\cite{absz95}. We also present the
Dalitz plot for the $\Upsilon(3S)\to\Upsilon(1S)\pi^+\pi^-$ decay
in Fig.~\ref{fig:dalitz}. It is shown that although the estimated
mass of $X$ in our model ($M_X =10.08$ GeV) is inside the data
area in the Dalitz plot, the signal of $X$ in the Dalitz plot is
not very clear due to its large width of 0.655 GeV. It does not
conflict with the CLEO experiment \cite{cleo98}. Moreover, we
would mention that with the typical values, $M_X$=10.5 GeV and
$\Gamma_X$=0.15 GeV given in Ref.~\cite{absz95}, we cannot produce
a $\cos\theta_\pi^*$ distribution that is consistent with the
experimental data\cite{cleo98}.

\section{Summary}

\label{sec:sum} Starting from an effective Lagrangian and further
employing ChUT to include the $\pi\pi$ $S$ wave FSI properly, the
$\pi^+\pi^-$ transitions of heavy quarkonia are intensively
studied. In order to consistently explain the $\pi^+\pi^-$
invariant mass spectra and angular distributions in the mentioned
processes simultaneously, especially in the
$\Upsilon(3S)\to\Upsilon(1S)\pi^+\pi^-$ decay process, an
additional sequential process, where an intermediate state $X$ is
introduced, is further considered in the bottomonium transitions.
With such a process included, all the $\pi^+\pi^-$ transition data
can be well-explained, especially the angular distribution of the
$\Upsilon(3S)\to\Upsilon(1S)\pi^+\pi^-$ decay. As a consequence,
the newly introduced intermediate state should have quantum
numbers of $J^P=1^+$ and $I=1$, a mass of about 10.08 GeV and a
width of about 0.655 GeV. The quark content of the state should be
$b\bar{b}q\bar{q}$. It might be a tetraquark state or a $B\bar{B}$
bound state. The detailed inner structure of the state should be
carefully studied both theoretically and experimentally.

\section*{Acknowledgements}

We would like to thank B.S. Zou for his valuable discussions and
suggestions. We are also benefit from fruitful discussions and
constructive comments given by E. Oset and D.O. Riska. We should
also thank F.A. Harris for providing us the BES data used in
\cite{bes00}. This project is partially supported by the NSFC
grant Nos. 90103020, 10475089, 10435080, 10447130 and CAS
Knowledge Innovation Key-Project grant No. KJCX2SWN02.

\end{document}